\pdfoutput = 1

\documentclass[12pt,a4paper]{article}

\usepackage{amsmath}
\usepackage{amssymb}

\usepackage[nosort,noadjust]{cite}

\usepackage{graphicx}

\begin{document}
\begin{titlepage}
\thispagestyle{empty}

\begin{flushright}
Nikhef-2018-059\\
\end{flushright}

\vspace{1.0truecm}
\begin{center}
\boldmath
\large\bf Probing New Physics in $B\to\pi K$ Decays
\unboldmath
\end{center}

\vspace{0.5truecm}
\begin{center}
 Robert Fleischer\,${}^{a, b}$, Ruben Jaarsma\,${}^{a}$ \footnote{Speaker}, Eleftheria Malami\,${}^a$  and  K. Keri Vos\,${}^{c}$\\[0.1cm]

${}^a${\sl Nikhef, Science Park 105, NL-1098 XG Amsterdam, Netherlands}

${}^b${\sl  Department of Physics and Astronomy, Faculty of Science, \\
Vrije Universiteit Amsterdam, NL-1081 HV Amsterdam, Netherlands}

${}^c${\sl Theoretische Physik 1, Naturwissenschaftlich-Technische Fakult\"at, \\
Universit\"at Siegen, D-57068 Siegen, Germany}
\end{center}

\vspace{1.0truecm}

\begin{center}
{\bf Abstract}
\end{center}

{\small
\vspace{0.2cm}\noindent
Over the last two decades, $B \to \pi K$ modes have received a lot of attention. They are particularly interesting since the main contribution to these decays comes from QCD penguin topologies. Furthermore, electroweak penguin amplitudes enter at a level comparable to the tree topologies. In the past, a discrepancy was found in the correlation between the CP asymmetries of $B_d^0 \to \pi^0 K_{\rm S}$. We provide the state-of-the-art picture of this situation and consider new constraints, finding that the effect has become larger. An attractive explanation is offered by a modified electroweak penguin sector. We propose a new method to determine the relevant parameters. It employs an isospin relation between the amplitudes of the $B \to \pi K$ decays, and makes only minimal use of the $SU(3)$ flavour symmetry. Using current data as well as future scenarios, we demonstrate how the mixing-induced CP asymmetry of $B_d^0 \to \pi^0 K_{\rm S}$ plays a key role in this analysis. The application of our strategy at the next generation of $B$-physics experiments may establish New Physics and reveal new sources of CP violation in the electroweak penguin sector.}

\vspace{1.0truecm}

\begin{center}
{\sl Invited talks at\\

\vspace{0.5\baselineskip}

The Sixth Annual Conference on Large Hadron Collider Physics (LHCP 2018)\\
Bologna, Italy, 4--9 June 2018\\

\vspace{0.25\baselineskip}

and\\

\vspace{0.25\baselineskip}

The Tenth International Workshop on the CKM Unitarity Triangle (CKM 2018)\\
Heidelberg, Germany, 17--21 September 2018\\

\vspace{0.5\baselineskip}

To appear in the Proceedings}
\end{center}

\vfill
\noindent
December 2018
\end{titlepage}

\thispagestyle{empty}
\vbox{}
\newpage

\setcounter{page}{1}

\section{Introduction}

The $B \to \pi K$ system has received a lot of attention over the last twenty years (\cite{BFRS-1,BFRS-2,FJPZ} and references therein). These modes are useful for studying electroweak (EW) penguin topologies, which may enter at a level comparable to the tree amplitudes; the CKM matrix element $V_{ub}$ suppresses the latter. Consequently, QCD penguin topologies give the leading contribution.

The $B_d^0 \to \pi^0 K_{\rm S}$ channel is of particular interest because it has a mixing-induced CP asymmetry. In the past, the correlation of this observable with the direct CP asymmetry has revealed an intriguing discrepancy. A modified EW penguin sector has been considered as an appealing explanation \cite{FJPZ}. We will first introduce the $B \to \pi K$ system, after which we provide a state-of-the-art analysis of the correlation between the CP asymmetries of the $B_d^0 \to \pi^0 K_{\rm S}$ mode. Finally, we propose a new strategy to determine the contribution of the EW penguin amplitudes \cite{FJV-1,FJMV}.

\boldmath
\section{The $B \to \pi K$ Decays}
\unboldmath

The four $B \to \pi K$ channels may receive contributions from tree, QCD penguin and EW penguin topologies. The first two enter the amplitudes through the hadronic parameters
\begin{equation} \label{eq:had-param}
	r_{\rm c}e^{i\delta_{\rm c}} \equiv (\hat{T}+\hat{C})/P', \quad re^{i\delta} \equiv (\hat{T}-\hat{P}_{tu})/P'.
\end{equation}
Here $\hat{T} (\hat{C})$ are colour-allowed (colour-suppressed) tree contributions, while $P' \propto P_{tc}$, where $P_{tc}$ denotes the difference between QCD penguins with internal $t$ and $c$ quarks; a similar comment applies to $\hat{P}_{tu}$. Using the $SU(3)$ flavour symmetry, these parameters can be determined from $B \to \pi \pi$ data, where EW penguin contributions are tiny \cite{BFRS-1,BFRS-2,FJPZ}. Allowing for non-factorizable $SU(3)$-breaking corrections of $20 \%$, we obtain \cite{FJV-1,FJMV}
\begin{equation} \label{eq:had-param-num}
	r_{\rm c}e^{i\delta_{\rm c}} = (0.17\pm0.06)e^{i(1.9\pm23.9)^\circ}, \quad re^{i\delta} = (0.09\pm0.03)e^{i(28.6\pm21.4)^\circ}.
\end{equation}
In a study of $B_{d,s} \to h^+ h^-$ decays $(h \in \{\pi,K\})$, we found no signs of anomalously large non-factorizable $SU(3)$-breaking effects \cite{FJV-2,FJV-3}.

The EW penguin topologies enter in colour-allowed and colour-suppressed form. The $B_d^0 \to \pi^- K^+$ and $B^+ \to \pi^+ K^0$ channels receive only contributions of the latter kind, which are expected to be small. However, the $B_d^0 \to \pi^0 K^0$ and $B^+ \to \pi^0 K^+$ decays receive contributions from both. These effects are described by the parameters \cite{NR-1,NR-2,BFRS-1,BFRS-2}:
\begin{equation} \label{eq:qphi}
	q e^{i\phi} e^{i\omega} \equiv - \left(\frac{\hat{P}_{EW} + \hat{P}_{EW}^{\rm C}}{\hat{T} +\hat{C}}\right) \stackrel{\rm{SM}}{=} \frac{-3}{2\lambda^2 R_b}\left(\frac{C_9 + C_{10}}{C_1 + C_2}\right) R_q = (0.68 \pm 0.05) R_q,
\end{equation}
which can be calculated in the SM using the $SU(3)$ flavour symmetry. Here, $\phi$ and $\omega$ are CP-violating and CP-conserving phases, respectively. Note that the latter parameter vanishes in the $SU(3)$ limit. It is actually a model-independent feature that $\omega$ is small \cite{BBNS}. Furthermore, $\hat{P}_{EW}^{(C)}$ are the colour-allowed (colour-suppressed) EW penguin topologies, the $C_i$ are short-distance coefficients, $\lambda \equiv |V_{us}| = 0.22$ is the Wolfenstein parameter, and $R_b$ is the usual side of the unitarity triangle (UT). The parameter $R_q = 1.0 \pm 0.3$ allows for $SU(3)$-breaking corrections. A precision of $5 \%$ appears achievable in the future thanks to expected progress in lattice QCD calculations \cite{FJPZ}.

Experimentally, we have access to branching ratios ${\cal B}^f$ and direct CP asymmetries $A_{\rm CP}^f$ for all four $B \to \pi K$ channels. Moreover, in the case of the $B_d^0 \to \pi^0 K_{\rm S}$ mode, we have also a mixing-induced CP asymmetry $S_{\rm CP}^f$. The direct CP asymmetries are all proportional to $r_{({\rm c})}\sin\delta_{({\rm c})}$. Consequently, they take values of at most $10 \%$ due to the smallness of these parameters, which are given in Eq.~(\ref{eq:had-param-num}). Furthermore, these observables enter the following sum rule \cite{G,GR,FJV-1,FJMV}:
\begin{align} \label{eq:sum-rule-theory}
\Delta_{\rm SR} &\equiv  
\left[A^{\pi^+ K^0}_{\rm CP}\frac{{\mathcal B}^{\pi^+ K^0}}{{\mathcal B}^{\pi^- K^+}}
-A^{\pi^0K^+}_{\rm CP}
\frac{2{\mathcal B}^{\pi^0 K^+}}{{\mathcal B}^{\pi^- K^+}}\right]\frac{\tau_{B_d}}{\tau_{B^\pm}}
+A^{\pi^-K^+}_{\rm CP} - A^{\pi^0K^0}_{\rm CP}\frac{2{\mathcal B}^{\pi^0K^0}}{{\mathcal B}^{\pi^-K^+}} = 0 + {\cal O}(r_{(c)}^2),
\end{align}
which is satisfied experimentally at the $1\sigma$ level \cite{PDG}. The $\pi^0 K^0$ final state is difficult to measure for LHCb. Consequently, $A_{\rm CP}^{\pi^0 K^0}$ has not received a lot of attention in recent years, and the current experimental average is \cite{PDG}
\begin{equation}
A_{\rm CP}^{\pi^0 K^0} = 0.00 \pm 0.13.
\end{equation}
This situation will improve with results from the Belle II experiment, where they aim to measure this observable with a precision of $4 \%$ \cite{Belle-II}. Using Eq.~(\ref{eq:sum-rule-theory}), we may actually predict this direct CP asymmetry from current data, yielding \cite{FJV-1,FJMV}
\begin{equation} \label{eq:aDir-sum-rule}
	A_{\rm CP}^{\pi^0K^0} = -0.14 \pm 0.03.
\end{equation}

The mixing-induced CP asymmetry enters the following time-dependent rate asymmetry:
\begin{equation}
	\frac{\Gamma(\bar{B}_d^0(t) \rightarrow \pi^0K_{\rm S}) - 
\Gamma(B_d^0(t) \rightarrow \pi^0K_{\rm S})}{\Gamma(\bar{B}_d^0(t) \rightarrow \pi^0K_{\rm S}) + 
\Gamma(B_d^0(t) \rightarrow \pi^0K_{\rm S})} = A^{\pi^0K_{\rm S}}_{\rm CP} \cos(\Delta M_dt) + S^{\pi^0K_{\rm S}}_{\rm CP}\sin(\Delta M_dt),
\end{equation}
where $\Delta M_d$ is the mass difference between the $B_d$ mass eigenstates. We can express the mixing-induced CP asymmetry in terms of $A^{\pi^0K_{\rm S}}_{\rm CP}$, as well as $\phi_{00} \equiv \rm{arg}(\bar{A}_{00} A^*_{00})$, the angle between the decay amplitude $A_{00} \equiv A(B_d^0 \to \pi^0 K^0)$ and its CP-conjugate $\bar{A}_{00}$ \cite{FJPZ}:
\begin{equation} \label{eq:corr-CP-asymmetries}
	S^{\pi^0K_{\rm S}}_{\rm CP} = \sin(\phi_d - \phi_{00})\sqrt{1- (A^{\pi^0K_{\rm S}}_{\rm CP})^2},
\end{equation}
where also the CP-violating $B_d^0$--$\bar{B}_d^0$ mixing phase $\phi_d=(43.2\pm1.8)^\circ$ enters. Using the hadronic parameters in Eq.~(\ref{eq:had-param}), we find \cite{FJV-1,FJMV}
\begin{equation} \label{eq:phi00}
	\tan\phi_{00} =  2(r\cos\delta-r_{\rm c}\cos\delta_{\rm c})\sin\gamma+2r_{\rm c} \left(\cos\delta_{\rm c}-2\tilde a_{\rm C} /3 \right) q\sin\phi +{\cal O}(r_{\rm (c)}^2).
\end{equation}
For completeness, we have included the contributions from colour-suppressed EW penguins through $\tilde a_{\rm C} \equiv a_{\rm C}\cos(\Delta_{\rm C} + \delta_{\rm c})$, even though they play a minor role \cite{BFRS-1,BFRS-2,FJV-1,FJMV}.

\boldmath
\section{The CP Asymmetries of $B_d^0 \to \pi^0 K_{\rm S}$}
\unboldmath

The $B \to \pi K$ amplitudes satisfy the following isospin relation \cite{BFRS-1,BFRS-2,FJPZ}:
\begin{equation} \label{eq:isospin-relation-amps}
	\sqrt{2} A(B^0_d \to \pi^0 K^0) + A(B^0_d \to \pi^- K^+) = \sqrt{2} A(B^+\to \pi^0 K^+) + A(B^+\to \pi^+ K^0) \equiv 3A_{3/2},
\end{equation}
where the isospin $I=3/2$ amplitude $3A_{3/2} \equiv 3 |A_{3/2}|e^{i\phi_{3/2}}$ is given by
\begin{equation} \label{eq:isospin-relation-qphi}
	3A_{3/2} = - (\hat{T} +\hat{C})e^{i\gamma} + (\hat{P}_{EW} + \hat{P}_{EW}^{\rm C}) = - (\hat{T} +\hat{C})\left(e^{i\gamma} - q e^{i\phi} e^{i\omega}\right),
\end{equation}
and $\gamma = (70\pm 7)^\circ$ denotes the usual UT angle. Applying once again the $SU(3)$ flavour symmetry, we obtain the relation \cite{GRL}
\begin{equation} \label{eq:tplusc}
	|\hat{T}+\hat{C}| = R_{T+C}\left|V_{us}/V_{ud}\right|\sqrt{2} |A(B^+\to\pi^+ \pi^0)|,
\end{equation}
where the $V_{ui}$ are elements of the CKM matrix. Within factorization, we have $R_{T+C} = f_K/f_\pi = 1.1928 \pm 0.0026$ \cite{PDG}. We will allow for non-factorizable $SU(3)$-breaking effects of up to $100 \%$ of the leading factorizable contribution, yielding $R_{T+C} = 1.2 \pm 0.2$ \cite{FJPZ,KMM}. Progress in lattice QCD calculations is expected to improve the precision with one order of magnitude \cite{FJPZ}.

The expression in Eq.~(\ref{eq:corr-CP-asymmetries}) describes a correlation between the CP asymmetries of $B_d^0 \to \pi^0 K_{\rm S}$. In order to utilize this relation, we require the angle $\phi_{00}$. It may be determined from Eq.~(\ref{eq:phi00}) using the values of the parameters in Eqs.~(\ref{eq:had-param-num})~and~(\ref{eq:qphi}). However, the cleanest way is to use the isospin relation in Eqs.~(\ref{eq:isospin-relation-amps})~and~(\ref{eq:isospin-relation-qphi}) \cite{FJPZ}. These expressions describe amplitude triangles in the complex plane, allowing the extraction of the angle $\phi_{00}$ from neutral $B \to \pi K$ decays. In this determination, no decay topologies have to be neglected, and the $SU(3)$ flavour symmetry enters only through $R_{T+C}$ in Eq.~(\ref{eq:tplusc}) and $R_q$ in Eq.~(\ref{eq:qphi}). Due to different triangle orientations, we encounter a fourfold ambiguity in the determination of $\phi_{00}$, which can be resolved through further considerations \cite{FJPZ,FJV-1,FJMV}.

\begin{figure}[t]
\begin{minipage}{0.5\linewidth}
\centerline{\includegraphics[width=0.66\linewidth]{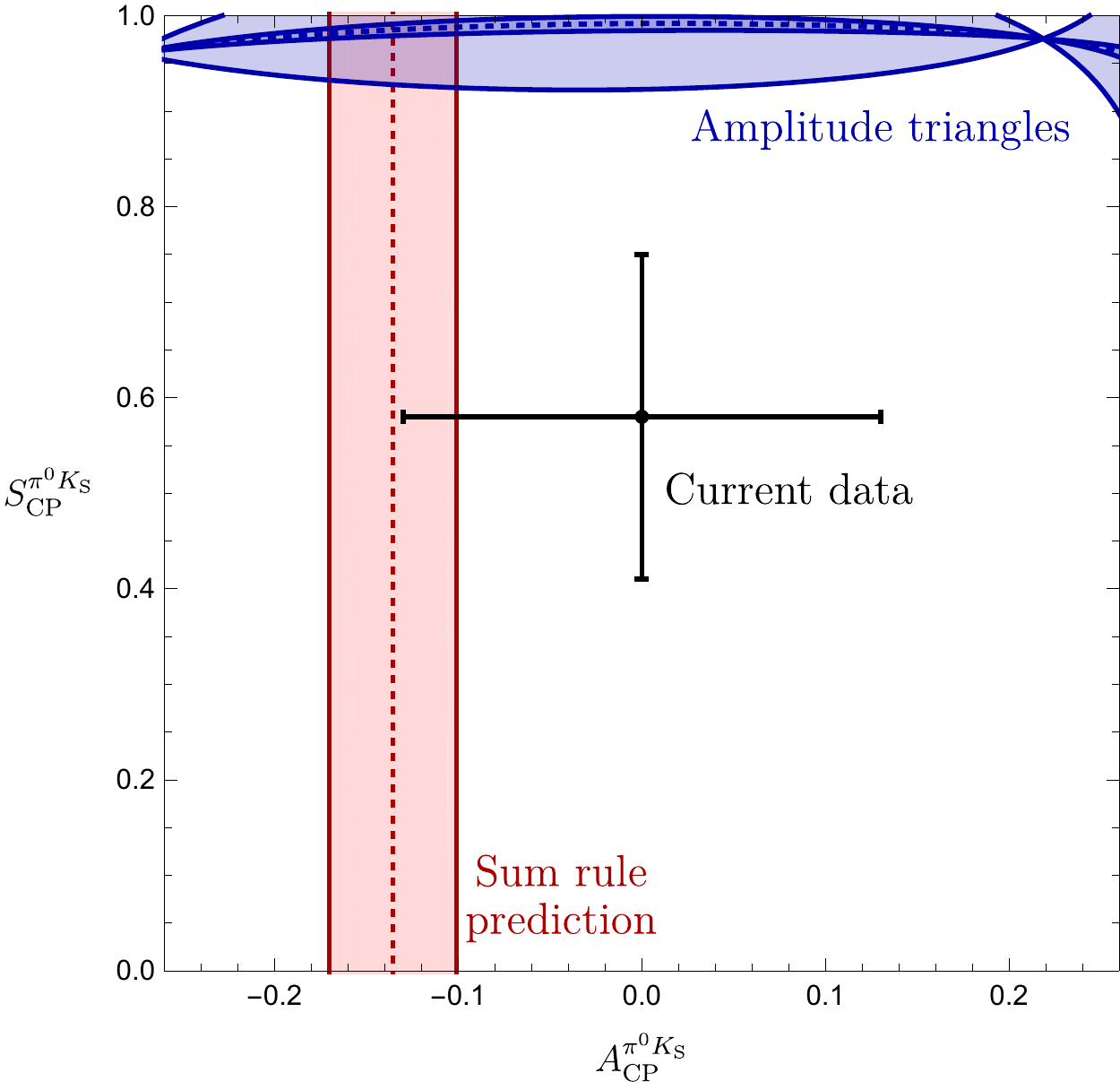}}
\end{minipage}
\hfill
\begin{minipage}{0.5\linewidth}
\centerline{\includegraphics[width=0.66\linewidth]{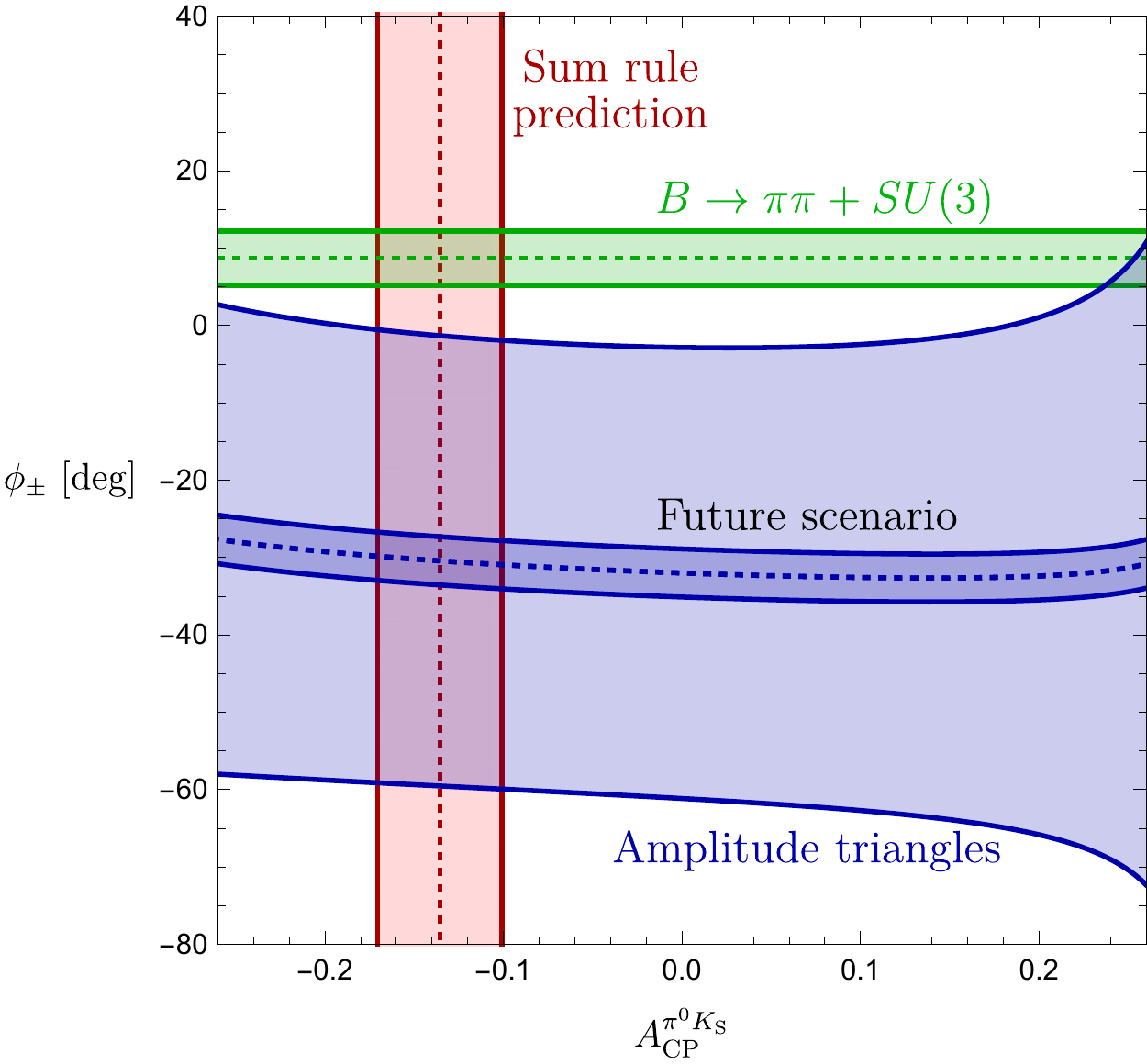}}
\end{minipage}
\caption[]{Left panel: Correlation between the CP asymmetries of $B_d^0 \to \pi^0 K_{\rm S}$ as derived from the amplitude triangles. Right panel: $\phi_\pm$ as a function of $A_{\rm CP}^{\pi^0 K_{\rm S}}$. Plots follow \cite{FJV-1,FJMV}.}
\label{fig:triangle-correlation}
\end{figure}

Finally, we find the correlation in the left panel of Fig.~\ref{fig:triangle-correlation} \cite{FJV-1,FJMV}. In comparison with previous work \cite{FJPZ}, the uncertainty of the contour has been significantly reduced, in particular due to more precise measurements of $\gamma$. The narrow band illustrates a future scenario, where we have assumed no experimental uncertainties, i.e.\ perfect measurements, as well as improvements in the determination of $R_q$ and $R_{T+C}$ due to progress in lattice QCD calculations, as discussed before. We have added a cross denoting the current experimental situation \cite{PDG}, and the red band corresponds to the prediction for $A_{\rm CP}^{\pi^0 K_{\rm S}}$ from the sum rule given in Eq.~(\ref{eq:aDir-sum-rule}). We note that there is a discrepancy between the contour from the amplitude triangles and the data at the $2.5~\sigma$ level.

An additional, new constraint can be obtained from the angle $\phi_\pm \equiv  \rm{arg}(\bar{A}_\pm A_\pm^*)$ between the decay amplitude $A_\pm \equiv A(B_d^0 \to \pi^- K^+)$ and its CP-conjugate $\bar{A}_\pm$, which can also be determined from the amplitude triangles. On the other hand, calculating this angle in terms of the hadronic parameters in Eq.~(\ref{eq:had-param}), we obtain the following simple expression for $\phi = 0^\circ$  \cite{FJV-1,FJMV}:
\begin{equation}
\left.\phi_\pm \right |_{\phi=0}= 2 \, r \cos\delta\sin\gamma + {\cal O}(r^2).
\end{equation}
Employing the numerical values in Eq.~(\ref{eq:had-param-num}), we find $\left.\phi_\pm \right |_{\phi=0} = (8.7 \pm 3.5)^\circ$, which yields the contour in the right panel of Fig.~\ref{fig:triangle-correlation}. We encounter again a discrepancy, which shows that the correlation from the amplitude triangles itself is in tension with the SM. The situation in Fig.~\ref{fig:triangle-correlation} could be resolved if the branching ratio of $B_d^0 \to \pi^0 K^0$ goes down by about $2.5\, \sigma$ while $S_{\rm CP}^{\pi^0 K_{\rm S}}$ increases by about $1\, \sigma$. On the other hand, this may also be a sign of NP. In this exciting case, a modified EW penguin sector offers an attractive scenario.

\section{Extracting the Electroweak Penguin Parameters}

\begin{figure}[t]
\begin{minipage}{0.32\linewidth}
\centerline{\includegraphics[width=\linewidth]{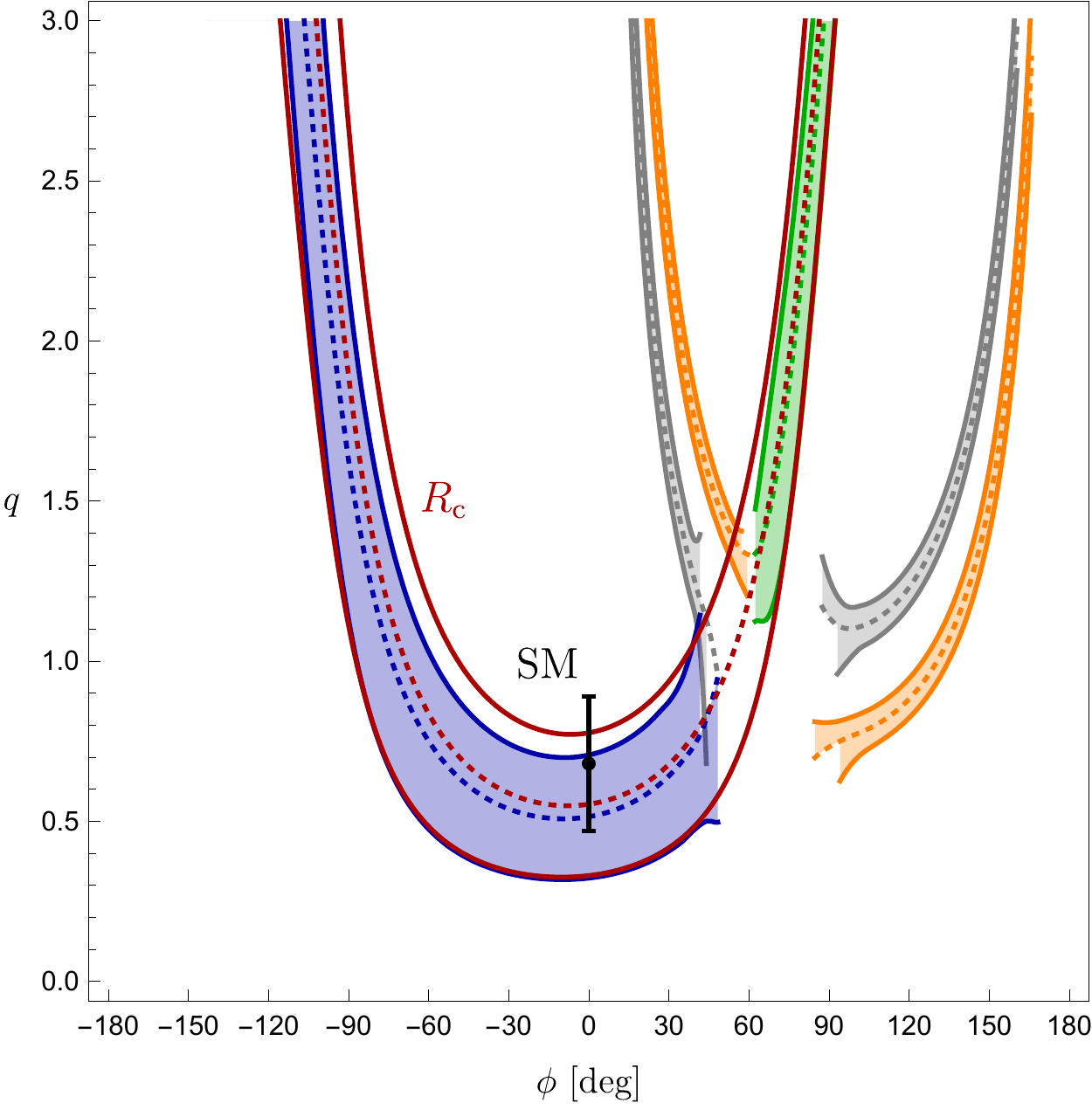}}
\end{minipage}
\hfill
\begin{minipage}{0.32\linewidth}
\centerline{\includegraphics[width=\linewidth]{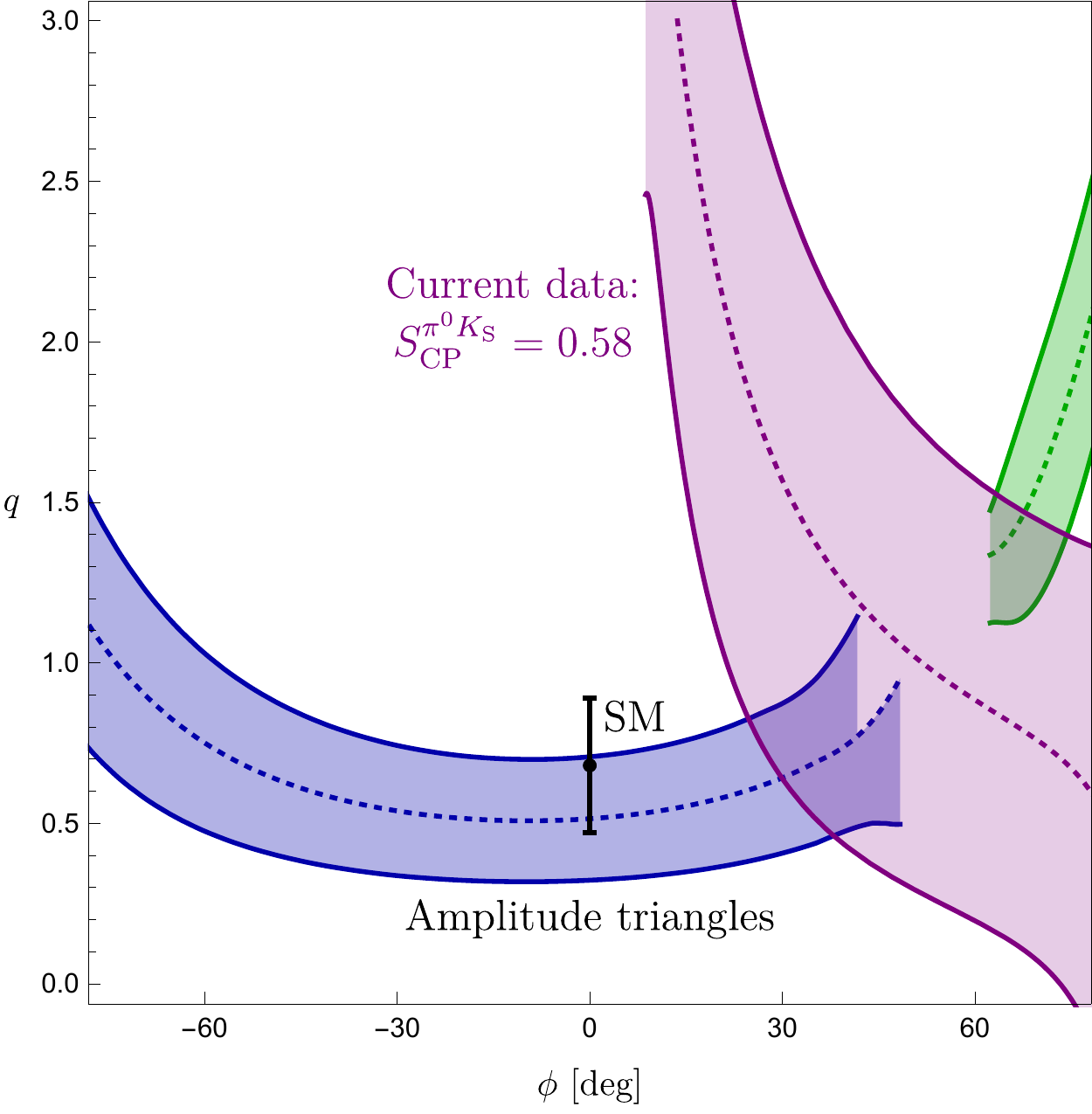}}
\end{minipage}
\hfill
\begin{minipage}{0.32\linewidth}
\centerline{\includegraphics[width=\linewidth]{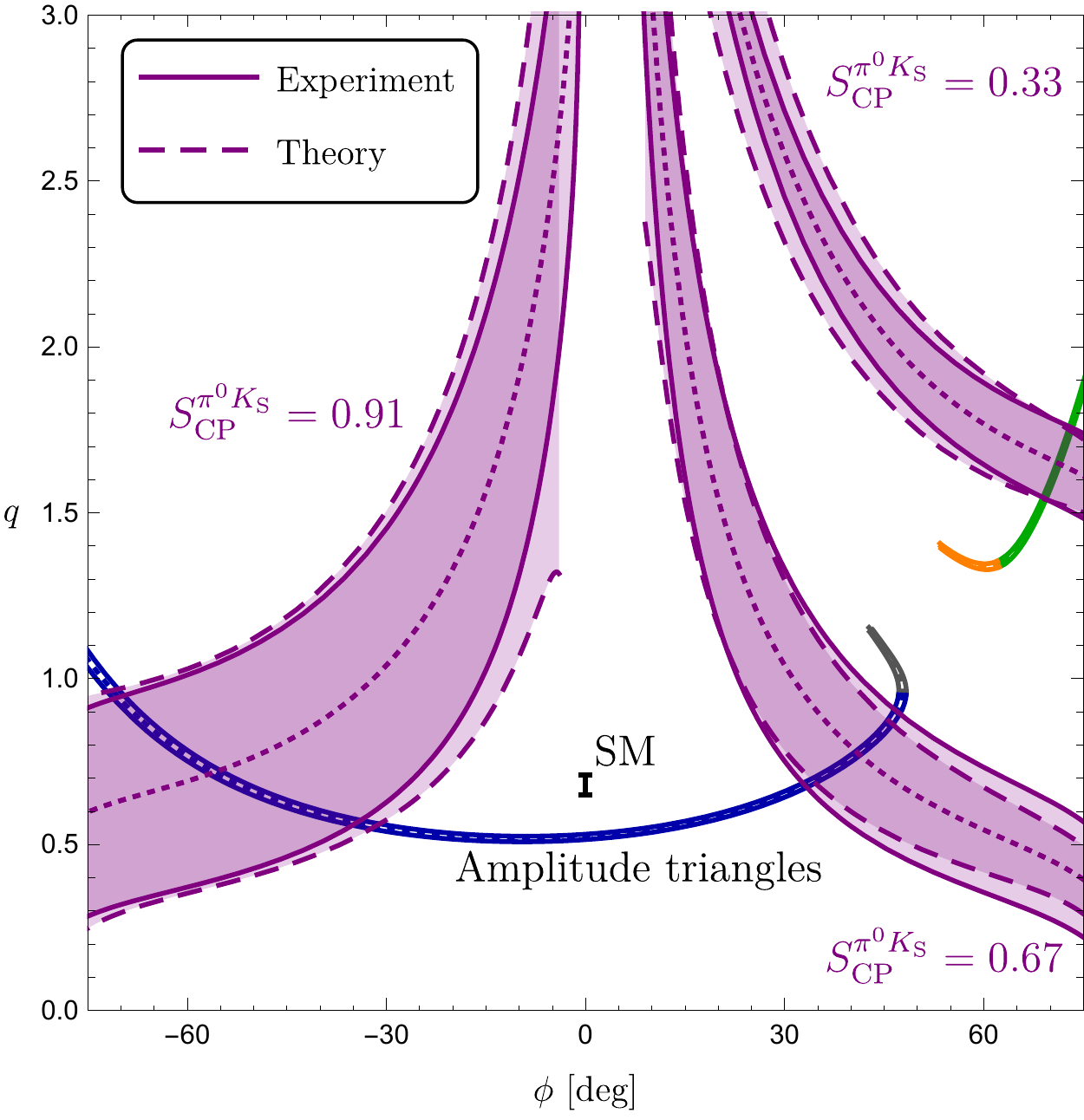}}
\end{minipage}
\caption[]{Contours in the $\phi$--$q$ plane from the amplitude triangles for charged $B \to \pi K$ data. In the left panel we show the situation for current data. The middle panel shows in addition the contour from $S_{\rm CP}^{\pi^0 K_{\rm S}}$, whereas the right panel illustrates future scenarios for measurements of this observable as discussed in the text. Plots follow \cite{FJV-1,FJMV}.}
\label{fig:qphi}
\end{figure}

The isospin relation in Eqs.~(\ref{eq:isospin-relation-amps})~and~(\ref{eq:isospin-relation-qphi}) can also be applied to put a constraint on the EW penguin parameters $q$ and $\phi$ \cite{FJV-1,FJMV}. This method can be separately implemented for the charged and neutral $B \to \pi K$ decays. It requires us to determine the relative orientation of the triangles. In case of the neutral decays, we can use $S_{\rm CP}^{\pi^0 K_{\rm S}}$ to determine $\phi_{00}$. As we have currently a large uncertainty of $S_{\rm CP}^{\pi^0 K_{\rm S}}$ \cite{PDG}, we will focus on the charged decays. In that case, we utilize the angle $\phi_{\rm c} \equiv \text{Arg}[\bar{A}_{+0}A_{+0}^\ast] = \mathcal{O}(1^\circ)$ between the decay amplitude $A(B^+ \to \pi^+ K^0)$ and its CP conjugate. We then obtain the contours in the left panel of Fig.~\ref{fig:qphi}, where we also show the SM expectation corresponding to Eq.~(\ref{eq:qphi}). It should be emphasized that no topologies have to be neglected to implement this method, and it requires only $SU(3)$ input to calculate $|\hat{T}+\hat{C}|$ in Eq.~(\ref{eq:tplusc}).

Further insights can be obtained from the ratio \cite{BFRS-1,BFRS-2,FJV-1,FJMV}
\begin{equation}
R_{\rm c} \equiv 2 \left[ \frac{{\mathcal B}(B^+ \to \pi^0 K^+)}{{\mathcal B}(B^+ \to \pi^+K^0)} \right] = 1-2 r_{\rm c}\cos\delta_{\rm c}(\cos\gamma-q\cos\phi)+{\cal O}(r_{\rm c}^2).
\end{equation}
Using current experimental data \cite{PDG}, we find $R_{\rm c} = 1.09 \pm 0.06$, yielding an additional constraint in the $\phi$--$q$ plane. It is shown as the red contour in the left panel of Fig.~\ref{fig:qphi}, and agrees remarkably well with the blue and green curves from the amplitude triangles.

In order to pin down the values of the EW penguin parameters we require further information. In particular, we may employ $S_{\rm CP}^{\pi^0 K_{\rm S}}$ to determine $\phi_{00}$. Eq.~(\ref{eq:phi00}) then yields an additional contour in the $\phi$--$q$ plane, where the numerical values of the hadronic parameters in Eq.~(\ref{eq:had-param-num}) are used as input. The expression is favourable from the theoretical side since the strong phases enter always with a cosine, giving only small variations with respect to these parameters. The contribution from colour-suppressed EW penguins, which enters Eq.~(\ref{eq:phi00}) through $\tilde a_{\rm C}$, can also be included using data. This is discussed in detail in Ref.~\cite{FJV-1,FJMV}. For current data, we then obtain the purple contour in the middle panel of Fig.~\ref{fig:qphi}. We show also the contours from the amplitude triangles in agreement with the $R_{\rm c}$ constraint, zooming in on the region around the SM point.

Additionally, we consider three scenarios for future measurements of $S_{\rm CP}^{\pi^0 K_{\rm S}}$, as given in the right panel of Fig.~\ref{fig:qphi}. For the contours from the amplitude triangles, we assume a more precisie determination of $R_{T+C}$ along with no experimental uncertainties, as for the future scenario in Fig.~\ref{fig:triangle-correlation}. For the CP asymmetries of $B_d^0 \to \pi^0 K_{\rm S}$, we have assumed an uncertainty of $\pm 0.04$, which can be reached at the end of Belle II \cite{Belle-II}. Moreover, we allow for non-factorizable $SU(3)$-breaking effects of $20 \%$ for the hadronic parameters. The experimental and theoretical error bands are shown separately in Fig.~\ref{fig:qphi} by the solid and dashed lines, respectively, illustrating that the experimental precision can be matched by theory.

\section{Conclusions}

Over the years, the $B \to \pi K$ data have shown puzzling patterns. We have presented a state-of-the-art analysis of the correlation between the CP asymmetries of $B_d^0 \to \pi^0 K_{\rm S}$, and found that a previous tension with the data has become stronger. In addition, we have considered a new constraint that shows that the correlation itself is not in agreement with the SM. To explain this pattern, either the data for $B_d^0 \to \pi^0 K_{\rm S}$ have to move to agree with the SM, or NP contributions are at work. In the latter case, a modified EW penguin sector is an interesting candidate.

We have presented a new strategy to determine the EW penguin parameters, which involves an isospin relation between the amplitudes of $B \to \pi K$ decays. Already for current data we obtain a very constrained situation. The analysis is complemented by the mixing-induced CP asymmetry of $B_d^0 \to \pi^0 K_{\rm S}$, allowing us to pin down the parameters describing the EW penguin contributions with unprecedented precision, which we have illustrated with three future scenarios. Consequently, the implementation of our strategy at future $B$-physics experiments may eventually establish New Physics with new sources of CP violation.


\section*{Acknowledgements}
This research has been supported by the Netherlands Organisation for Scientific Research (NWO) and by the Deutsche Forschungsgemeinschaft (DFG), research unit FOR 1873 (QFET).


%
%
%

%
%
%

\end{document}